\documentstyle[psfig,conf_iap,10pt]{article}
\begin{document}

\heading{Structure and Evolution of The Intergalactic Medium: \\
 Conference Summary  } 

\par\medskip\noindent
\author{ J. Michael Shull }

\address{Department of Astrophysical \& Planetary Sciences, 
CASA and JILA, \\ 
Campus Box 391, University of Colorado, Boulder CO 80309 (USA)  }

\begin{abstract}
Wasn't this a fun meeting?  Yes, except for the rain.
This summary highlights four scientific themes of the $13^{\rm th}$ IAP
conference, plus a ``wishlist'' of future projects for 
observers and theorists. 
\end{abstract}

\section{Introduction:  A Privileged Era? }
  
In agreeing to summarize the results of this meeting, I had
a few moments of doubt.  First, we were all dismayed not to
benefit from John Bahcall's 30 years of wisdom in the
field.  Second, a responsible effort would require 
many hours of close attention to 4.5 days of fascinating talks,
diverting time from beautiful walks and museums.  Most of all, 
I had a lingering concern that, on the last day of the conference, 
Jerry Ostriker would arrive from Princeton, just in
time to set us straight! As it turned out, this was an enjoyable task,
with stimulating talks on new data and new ideas.  

I believe our field of QSO absorption-line studies is 
in a privileged era.   Like the Roman god Janus, we look
backward toward the past and forward to the future, both in our 
scientific tools and our theoretical paradigms (Table 1).     
 
\begin{center}
\begin{tabular}{l c l}
\multicolumn{3}{l}{{\bf Table 1.} Examples of ``Janus-Like Era''} \\
~   \\
\hline
\multicolumn{1}{c}{\bf The Past} & 
    \multicolumn{1}{c}{$\longleftarrow \; | \; \longrightarrow$} &
     \multicolumn{1}{c}{\bf The Future} \\
\hline 
                            &              &                         \\
    4-meter telescopes      & ..........   & 8-10 meter telescopes   \\
    Galactic halos          & ..........   & CDM/hydro paradigm      \\
    Interstellar models     & ..........   & Cosmological models     \\
    Ly$\alpha$ clouds + IGM & ..........   & The ``cosmic web''      \\ 
\hline
\end{tabular}
\end{center}

\vspace{0.1cm}

Since the discovery of the high-redshift Ly$\alpha$ forest over
25 years ago, these absorption features in the spectra
of QSOs have been used as evolutionary probes of the intergalactic
medium (IGM), galactic halos, and now large-scale structure
and chemical evolution.  It is fascinating how rapidly our 
interpretation of these absorbers has changed, since 
they were interpreted as relatively small (10 kpc),
pressure-confined clouds of zero-metallicity gas left 
over from the era of recombination.  To be sure, the lack of
strong clustering in velocity provided ample grounds for 
distinctions from QSO metal-line systems and galactic halos.
However, these distinctions are clearly weakening.  

In the next few years, I expect that many of the 
divisions between research in ``interstellar''
and ``intergalactic'' matter and between ``QSO absorption clouds'' and
``cosmological structure'' will fade away.  We may even 
begin to understand more about how galaxies and their halos were
assembled.  Replacing the individual area studies will be a number
of hybrid problems: 
\begin{itemize}

\item CDM/Hydrodynamics + Feedback from star formation

\item Interface between galaxies, the ISM, and the IGM

\item Chemical evolution and heavy-element transport  

\item Reionization and the assembly of galaxies

\end{itemize}
Substantial portions of our July 1997  meeting were spent discussing
these issues.  Because I cannot do justice 
to all the individual talks (65 by my count), I will instead 
describe several outstanding problems in four scientific areas:
(1) {\it The History of Baryons}; (2) {\it The History of Metals};
(3) {\it Reionization of the IGM}; and (4) {\it The Assembly of Galaxies}.  
I will conclude by providing ``wish lists'' of scientific projects  
for observers, instrumentalists, and theorists.
 
\section{Major Themes of the Conference}

\subsection{The History of Baryons }

One of the compelling reasons to study intergalactic
Ly$\alpha$ clouds is that they may contain an appreciable fraction
of the high-$z$ baryons.  To the extent that the Ly$\alpha$ 
absorbers are associated with large-scale structure and galaxy
formation, the evolution of the IGM should parallel the evolution
of galaxies and the history of baryons.     
Therefore, a major task is to understand the 
physical significance of various features in the column-density
distribution of Ly$\alpha$ absorbers.  As shown in Figure 1, 
\begin{figure}
    \centerline{\vbox{
    \psfig{figure=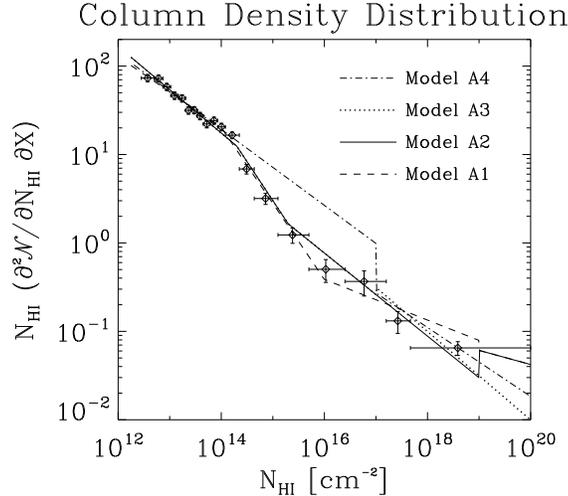,height=7cm} }}
     \caption[]{Column density distribution of Ly$\alpha$ clouds 
      at mean redshift $\langle z \rangle = 3$, based on {\it Keck} 
      and other spectra (see Fardal, Giroux, \& Shull 1997).  
      The variable $dX = (1+z) dz$ is the ``absorption length'' 
      (Petitjean et al.  1993).  Model A2 is a preferred fit
       with a turnover above $10^{14.5}$ cm$^{-2}$.} 
\end{figure}
Ly$\alpha$ absorbers range over nearly 10 orders of magnitude
in H~I column density, roughly from $10^{12}$ cm$^{-2}$ to 
$10^{22}$ cm$^{-2}$.  At the lower end, the {\it Keck} Telescope 
has detected weak Ly$\alpha$ absorption down to 
log N$_{\rm HI} \approx 12.3$. At the upper end, damped Ly$\alpha$
absorbers have been seen up to log N$_{\rm HI} \approx 21.6$.  
What are the physical reasons for this range and for 
features in the approximate power-law distribution? More
specifically,  we should be concerned about the following 
questions and issues:
\begin{itemize}

\item Is there a turnover in the distribution at log N$_{\rm HI} \leq 12.7$?
    These weak absorbers, which may arise in very low-density regions of
    the IGM, may produce substantial He~II absorption toward 
    high-$z$ QSOs.   

\item What is the physical significance of the steepening in the
    distribution above N$_{\rm HI} = 10^{14.5}$ cm$^{-2}$?  
    This turnover has been noticed for years, but it was difficult to verify 
    owing to curve-of-growth uncertainties.  Clouds at
    $10^{14-15}$ cm$^{-2}$ may contain most of the baryons 
    in the Ly$\alpha$ forest (for a $N_{\rm HI}^{-1.5}$ distribution).  
    Clouds at $10^{14.5-15.5}$ cm$^{-2}$ are used 
    for double-QSO cloud size estimates and for metal-line 
    detections of C~IV and Si~IV.  We need to understand their
    structure and shape. 

\item Can we detect the transition from atomic (H~I) to molecular
    (H$_2$) gas in the damped Ly$\alpha$ absorbers?  The expected
    turnover should be seen above log N$_{\rm HI} \approx 21.5$
    and related to high-redshift CO and the first stars.

 \item  For chemical evolution models, it is important to reconcile the 
    baryon evolution rate, $\dot{\Omega_b}$, with the star formation rate, 
    $\dot{\Omega_*}$ and the metal formation rate, $\dot{Z}$. 
    What is the role of the IGM in this network?  The Ly$\alpha$ forest 
    probably contains substantially more baryons than the damped Ly$\alpha$
    absorbers ($\Omega_{\rm LF} \approx 10 \, \Omega_{\rm DLA}$).  
    The metallicity of the Ly$\alpha$ forest is 
    $10^{-2.5 \pm 0.5}$ solar, while that of the
    damped Ly$\alpha$ absorbers is $10^{-1.5 \pm 0.5}$ solar. 
    The larger baryon reservoir in the forest may therefore be a significant 
    part of the metal inventory.  

\end{itemize} 

\subsection{The History of Metals }

Five years ago, most astronomers believed the Ly$\alpha$ forest clouds 
to be pristine.  The observations that a high percentage of 
Ly$\alpha$ clouds with N$_{\rm HI} \geq 10^{14.5}$ cm$^{-2}$ contain heavy 
elements (C~IV, Si~IV) were astonishing. Recent estimates of the metal 
abundance are $10^{-3.0}$ to $10^{-2.5}$ times solar metallicity and 
suggest a (Si/C) enhancement by about a factor 2 over solar ratios. 
The Si~IV lines are especially interesting,
since Si is thought to be formed by $\alpha$-capture processes
in massive stars and expelled by Type~II supernovae.  
We need to clarify some implications of these data:
\begin{itemize}

\item Where and when were the heavy elements formed? 
    Although the {\it Hubble} Deep Field and related observations 
    suggest that the bulk of metal production and star formation
    occurred  at $z = 1-2$, the metals in the Ly$\alpha$ forest 
    obviously formed earlier. 
    How much earlier?  Was it in disks, dwarf galaxies, or low-mass
    objects such as proto-globular clusters?  Were these metals
    blown out, stripped by mergers, or transported by other means? 
    Understanding these processes may clarify the ``astration'' 
    of deuterium by the first generations of stars.

\item Wherever the heavy elements were produced,  
    they cannot have been transported far from their source.  Over
    1 Gyr, metal-laden gas moving at 100 km~s$^{-1}$ would travel 
    100 kpc, approximately the size of typical galactic halos and some 
    Ly$\alpha$ clouds.  Local effects from massive star formation  
    could cause the ionization states of Si and C to differ significantly 
    from pure QSO photoionization.  A key experiment would be to
    infer the size of the metal-bearing Ly$\alpha$ clouds from 
    double-quasar coincidences. However, these moderate-column 
    (log N$_{\rm HI} > 14.5$) clouds 
    are sufficiently rare that good statistics will be difficult to obtain. 

\item Converting the observed N(Si~IV)/N(C~IV) to accurate Si/C abundances
     requires a clear understanding of the ionization mechanism 
     (photoionization versus collisional ionization).  If photoionization,
     we need a much better idea of the spectral shape, $J_{\nu}(z)$, 
     produced by quasars and starburst galaxies at redshifts $z = 2-5$. The 
     photon range from 2--5 Rydbergs is particularly important,
     since it covers the ionization edges of relevant Si and C ions.  
     In the Ly$\alpha$ forest, it is important to understand
     the Si~IV/C~IV ratios, component by component, as is often done in
     the analysis of interstellar absorption profiles.

\item  The abundances of the iron-group and other elements (Fe, Ni, 
     Zn, Si, Cr) in the damped Ly$\alpha$ systems need to be
     confirmed.  Their ratios provide strong suggestions of massive-star
     nucleosynthesis and hints of dust depletion.  Zinc may provide 
     particularly important clues. 

\item I was impressed by attempts to invert 
    the Mg~II, Fe~II, and C~IV line profiles to produce kinematics of 
    the metal-line absorbers.  However, I suspect that in most cases 
    the inversion is not unique; departures from simple orbits are
    likely, as shocks and gas dynamics are important.   

\end{itemize}

\subsection{Reionization of the IGM}

Most of us carry a mental picture that the IGM was reionized at
high redshift ($z \geq 5$) by the first quasars and first massive stars.
There are hints that it could occur even earlier.  
However, the redshift history of reionization is poorly known,
except for theoretical prejudices based on CDM models for structure
formation. Up to $z \approx 4$, the quasar luminosity function
is fairly well known, but the same cannot be said of quasars
at $z > 4$ or of starburst galaxies at any redshift.  
The following issues remain controversial:
\begin{itemize}

\item Are there missing QSOs at $z > 4$ owing to dust obscuration?
    Theoretical suggestions (Fall \& Pei 1989) of a substantial population 
    of ``missing quasars'' have not yet been confirmed. There are hints of 
    dust depletion from Zn/Cr ratios in damped Ly$\alpha$ systems, and
    conflicting results from red- and radio-selected high-$z$ quasars.
    One recent QSO luminosity function (Pei 1995) produces
    too few Lyman continuum photons to reionize hydrogen by $z = 4$.
    Thus, additional ionizing sources are needed:  either QSOs or
    starburst galaxies.

\item When were the {\it first} O stars?  If massive star formation
    is a natural result of the first star formation, then these
    objects will dominate the feedback to the gaseous environment,
    including dissociation of H$_2$, production of key heavy elements
    (O, Si, S), and generation of large volumes of hot gas
    through supernovae and stellar winds.  The details of this
    feedback depend on the galactic environment (dwarfs, spirals, halos).   

\item How much of the energy input to the IGM is mechanical?
    Massive stars produce both hot gas and ionizing radiation.
    If the ``mechanical'' energy input is released through blast
    waves, the affected volume scales with energy as as $E^{3/5}$,
    whereas metal production scales as $E$.  Thus, low-luminosity sources 
    (dwarf galaxies) may dominate metal dissemination.  
      
\item Have we detected the era of helium reionization?
    Dieter Reimers showed us intriguing evidence for patchy He~II absorption
    toward a quasar at $z = 2.9$. Can this be reconciled
    with Gunn-Peterson observations at $z \approx 4.7$ that suggest 
    reionization in hydrogen?  Theoretical models of the
    QSO luminosity function and IGM opacity suggest  
    that the IGM should be reionized in He~II at $z \geq 3.3$.  
    Perhaps this is evidence for hot-star ionizing sources, 
    with little 4 Ryd (He~II) continuum.

\end{itemize}

Observations are badly needed to push our understanding of
the reionization epoch back to $z \geq 5$.  First, we need to
find QSOs at $z > 4.9$, perhaps from the Sloan Sky Survey. 
Possible probes of the high-$z$ era include searches in the 
radio, microwave, far-infrared,
and near-infrared bands. As planning begins for the NGST 
(Next Generation Space Telescope), the infrared band
($1-5~\mu$m) offers promise for deep searches for high-$z$, 
dust-obscured QSO as well as high-$z$ supernovae.  
Detecting 21-cm emission at $z > 5$ might be possible with a 
sufficiently large array of
radio dishes.  The Ly$\alpha$ absorption from the neutral
IGM prior to reionization might show up in high-$z$ spectra of
quasars at $z \approx 5$.  To probe even higher redshifts, one
might consider searches for redshifted metal fine-structure lines
such as [C~II] 158$\mu$m and [O~I] 63$\mu$m, which would appear at  
$(1.6~{\rm mm})[(1+z)/10]$ and $(630~\mu{\rm m})[(1+z)/10]$.

\subsection{The Assembly of Galaxies }

The standard (CDM) model of galaxy formation predicts a
``bottom-up'' hierarchy of structure formation.  If clumps of
$10^{5-7}~M_{\odot}$ form massive stars 
at $z > 3$, they could have significant
effects on Lyman continuum radiation, hot gas, and heavy-element transport.  
If the sub-clumps form in the halos of proto-galaxies, or
fall in gravitationally, cloud-cloud
collisions are likely to occur. What are the implications of the
resulting shock waves for line profiles of Mg~II and C~IV
absorbers?  Shocks will generate hot gas at
$T \approx (10^5~{\rm K})[V/100~{\rm km~s}^{-1}]^2$,
sufficient to produce C~IV by collisional ionization.
Are the observed line profiles evidence for such effects?

Finally, we heard several speakers speculate on the formation
of large gas disks in the context of damped Ly$\alpha$
absorbers.  Are these DLAs actually thick disks of 30~kpc size 
or 5--8 kpc clumps as predicted by some numerical modelers?  
If the DLAs are as small as 5--8 kpc, it may be difficult to
understand their frequency, $d{\cal N}/dz$, and there may
be an angular momentum problem.  Following the
implications of the CDM scenario, how are the small pieces
of proto-galaxies assembled?  What are the roles of radiative
cooling and sub-clump mergers?  

\section{A Wish-List for the Future}

I conclude this review with lists of
ideas and scientific tools for workers in our field.  
I have given separate discussions for observers and theorists.

\subsection{Observers and Instrumentalists } 

I continue to be amazed by the beauty of the HIRES
spectra taken by the {\it Keck Telescope}.  These new optical
data have changed the field of QSO absorption lines in so
many areas.  My first wish is that the new 8--10$^m$ telescopes
and spectrographs become sufficiently productive to compete
with {\it Keck}.  Even though many astronomers are actively
using {\it Keck} for QSO studies, we can foresee the
time when several new telescopes come on line:
the {\it Hobby-Eberly Telescope}, the {\it VLT}, and {\it Gemini}.  

A general lesson learned from the {\it Keck}
experience is that high-resolution spectroscopy is one of the
most powerful tools in astrophysics.  That power should be extended 
to other wavelength bands.  Large instruments are needed:  
\begin{itemize}

\item {\bf Ultraviolet:}  To study D/H evolution, the
   He~II Gunn-Peterson effect, chemical evolution of metals,
   and damped Ly$\alpha$ systems, we need a UV spectrograph
   with effective area $\sim10^4$ cm$^2$.  For comparison,
   the current spectrographs aboard {\it Hubble} have
   $A_{\rm eff} \approx 100-200$ cm$^2$ for moderate-resolution
   (30-50 km~s$^{-1}$).  In 2002, HST will be upgraded with the
   {\it Cosmic Origins Spectrograph}, which will provide
   1000-1500 cm$^2$ effective area.  The next generation of
   UV instruments should consider taking another factor-of-ten
   step in spectroscopic throughput.

\item {\bf Infrared \& Sub-millimeter:}  Both the NGST and
   FIRST telescopes are designed to provide 4-meter apertures
   that access the near-IR and sub-mm respectively.  These
   instruments should provide powerful imaging of the era
   beyond redshift $z = 5$.  The spectroscopic capabilities
   may provide further surprises for detecting protogalaxies
   along with their first stars and supernovae.

\item {\bf Millimeter:}  As noted earlier, relating the damped
   Ly$\alpha$ absorbers (proto-galactic gaseous disks) to the
   first spiral galaxies will require us to follow the transition
   from atomic to molecular gas (from H~I to CO). The mm-array
   offers the chance to make these comparisons.

\item {\bf X-Ray:}  The equivalent X-ray instrument for high-resolution
   spectroscopy with sufficient throughput to match the optical 
   could be the HTXS (``High Throughput X-Ray Spectroscopy'') 
   mission.  Designed with 1--10 m$^2$ of effective area, this
   set of X-ray telescopes would have the capability of studying
   the ``X-ray Gunn-Peterson effect'' in heavy-element K-edge
   absorption through the metal-contaminated parts of the IGM. 
   These observations could detect the hot gas invisible 
   in H~I and He~II absorption.  These absorption signatures are
   expected to be extremely weak ($\tau \approx 10^{-3}$).

\end{itemize}

\subsection{Simulations and Modelers }

Many of the talks at this meeting worked within the new
cosmological paradigm for the Ly$\alpha$ clouds.  Over the
past five years, numerical models have increased their
accuracy and predictive power immensely, to the point where
they are now able to provide constraints on $\Omega_b$,
$J_{\nu}$, and galaxy formation.  There is still some ways
to go, however, and I offer the following wishes:
\begin{itemize}

\item Computers are increasing in their speed and capacity.
   Many of us, including the modelers, anticipate seeing their 
   simulations run to $z \rightarrow 0$ and computed in a
   box of size $100h^{-1}$ Mpc.  

\item The models need a better justification for the redshift
    at which quasars and star formation turn on.  As noted earlier 
    (\S2.3) we have little information on these epochs of reionization.

\item The models need to incorporate better small-scale physics.
    I have the impression that the gravitational collapse of large-scale
    gaseous structures is treated fairly well.  However, once stars and 
    quasars turn on, the microphysics [supernovae, stellar winds, 
    superbubbles, heavy element transport, hot gas, radiative transfer] 
    needs to be handled in a realistic fashion.  These ``local
    effects'' are the next hurdle in complexity.

\item To agree with the H~I column density distribution, the 
    numerical models require a lower ionizing radiation 
    field than that inferred from the proximity effect.  
    Models for the Ly$\alpha$ absorber distribution constrain
    the ratio $(\Omega_b^2 / J_{-21})$, where $J_{-21}$ is the
    specific intensity at the Lyman limit, in units of $10^{-21}$
    ergs cm$^{-2}$ s$^{-1}$ Hz$^{-1}$ sr$^{-1}$.  These parameters 
    need to be reconciled with independent inferences of the
    baryon density from deuterium measurements, $\Omega_b h^2 \approx 
    0.020 \pm 0.015$ (Tytler 1997), and of the radiation field 
    $J_{-21} \approx 0.5-1.0$ (Giallongo et al. 1996; Cooke et al.
    1997).  Some of the models described at this meeting suggest 
    a sizeable discrepancy. For example, Zhang et al. (1997)
    require a photoionization rate $\Gamma_{\rm HI}  = 
    (3-10) \times 10^{-13}$ s$^{-1}$, which corresponds to
    $J_{-21} \approx 0.1 - 0.4$ for background spectral slope
    $\alpha_b \approx 1.8$.   
     
\item Fluctuations in the ionizing background are quite important and
      should be included in the models.  The patchy He~II absorption
      reported by Reimers may be one manifestation of this.
      More generally, the baryons are not exposed to a constant,
      optically-thin radiation field.  

\end{itemize}

\acknowledgements{
On behalf of all of us attending this conference, 
I would like to express my thanks to the conference organizers and
to the IAP for their hospitality. Special appreciation is due to
Patrick Peitijean, Patrick Boiss\'e, Francoise Combes, St\'ephane 
Charlot, and Brigitte Raban. 
}

\vfill
\end{document}